\documentclass[11pt, titlepage]{article}
\usepackage{latexsym}
\usepackage{amsmath}
\usepackage{amssymb}
\usepackage{epsfig}
\usepackage{rotating}
\usepackage{multirow}
\usepackage{setspace}
\usepackage{graphicx}
\makeatletter
\renewcommand{\@seccntformat}[1]{{\csname the#1\endcsname}.\hspace{0.5em}}

\usepackage{harvard}

\graphicspath{{IMAGE_FILES/}}

\title{Dominance, Intimidation, and `Choking' on the PGA Tour\thanks{Robert A. Connolly is Associate Professor, Kenan-Flagler Business School, University of North Carolina, Chapel Hill.  Richard J. Rendleman, Jr. is Visiting Professor, Tuck School of Business at Dartmouth and Professor Emeritus, Kenan-Flagler Business School, University of North Carolina, Chapel Hill. The authors gratefully acknowledge the assistance of Carl Ackermann and the comments of Mark Broadie, Bill Perreault and participants in the Applied Statistics seminar at Dartmouth College, especially Michael Herron, Richard Sansing, Matt Slaughter, Paul Wolfson and Kent Womack.  Connolly gratefully acknowledges research support from the Richard "Dick" Levin Faculty Excellence Fund. Please address comments to Robert Connolly (email: Robert\_Connolly@unc.edu; phone: (919) 962-0053) or to Richard J. Rendleman, Jr. (e-mail: Richard\_Rendleman@unc.edu; phone: (919) 962-3188). } }

\author{Robert A. Connolly and Richard J. Rendleman, Jr.}

\date{This version: September 22, 2008}

\begin{document}
\bibliographystyle{jf}

\maketitle

\newpage
\thispagestyle{empty}

\begin{center}
{ \Large {How Dominant and Intimidating is Tiger Woods and \\ Does Phil Mickelson Really Choke?}

 \vspace{0.25in}

 {Abstract}
}
\end{center}

Extending the work of Connolly and Rendleman (2008), we document the dominance of Tiger Woods during the 1998-2001 PGA Tour seasons.  We show that by playing ``average,'' Woods could have won some tournaments and placed no worse than fourth in the tournaments in which he participated in year 2000, his best on the PGA Tour.  No other PGA Tour player in our sample could have come close to such a feat.  We also are able to quantify the intimidation factor associated with playing with Woods.  On average, players who were paired with Woods during the 1998-2001 period scored 0.462 strokes per round worse than normal.  Although we find that Woods' presence in a tournament may have had a small, but statistically significant adverse impact on the entire field, this effect was swamped by the apparent intimidation factor associated with having to play with Tiger side-by-side.

We also demonstrate that Phil Mickelson's performance in major golf championships over the 1998-2001 period was not nearly as bad as was frequently mentioned in the golf press.  Although Mickelson won no majors during this period, he played sufficiently well to have won one or two majors under normal circumstances.  Moreover, his overall performance in majors, relative to his estimated skill level, was comparable to that of Tiger Woods, who won five of 16 major golf championships during our four-year sample period.  Thus, the general characterization of Woods as golf's dominant player over the 1998-2001 period was accurate, but the frequent characterization of Phil Mickelson choking in majors was not.

\vspace{.2in}
\noindent
\newpage

\setcounter{page}{1}

\doublespacing
\section{Introduction}

\subsection{Overview}

By the spring of 2000, after having played on the PGA Tour for less than four years, Tiger Woods had become well established as golf's most feared and intimidating competitor.   Typifying this perception, the following appeared on the Golf Today website immediately before the 2000 Masters:
\begin{quote}
\singlespacing
``Tiger Woods, in the midst of crafting one of golf's most dominating stretches of brilliance and the runaway favorite at the Masters, is becoming the Intimidator of world golf.  Seeing Woods atop a Sunday leaderboard has signified a fight for second place among the rest of the world's best. The last 13 times he held the lead or was tied for the lead going into the final round he has closed the deal.  Woods bearing down on a front-runner has led to gruesome breakdowns and embarrassing collapses sometimes uncomfortable to watch on video replay.  The winner of seven of his last 11 tournaments and second or tied for second in three of the others does not deny the effect his hypnotic form has had on some rivals.'' (Golf Today, 2000).

\end{quote}

\doublespacing
\noindent
Consistent with this perception,  Gladwell  (2008) states:
\begin{quote}
\singlespacing
``We've seen it a thousand times;  top-notch professional golfers crumble on Sunday when in the final pairing with Tiger Woods. Why does this happen with such regularity? What is it that Tiger Woods does to intimidate his fellow golfers and almost guarantee victory for himself?''
\end{quote}

\doublespacing

In stark contrast, prior to his winning the 2004 Masters, and as of this writing, winning two more of golf's ``major'' championships, Phil Mickelson had been labeled as ``the best player in the world to have never won a major.''\footnote{Golf's majors include The Masters, The U.S. Open, The British Open and The PGA Championship.} \footnote{A Google search on the combined strings ``phil mickelson'' and ``to have never won a major'' brings up 1,310 references but, of course, there are other forms that the expression ``to have never won a major'' can take, which are not part of the search.}   Although he had become one of golf's premiere players, he was often accused of choking in majors, especially in the final rounds.  Describing Mickelson's propensity to choke, Smith (2001) wrote the following immediately prior to the 2001 PGA Championship:

\begin{quote}
\singlespacing
Mickelson will be among the favourites again, especially on an Atlanta Athletic Club course with the kind of length -- 7,213 yards for a par 70 -- that suits big hitters, and rain-softened greens that will allow him to attack the pins. Then again, his aggressive style is what has cost him so many chances. Mickelson challenges just about every flag, as if the thought never crosses his mind that he might hit a bad shot. The result is missing the green on the short side, leading to bogey or worse. Alas, what Mickelson lacks in majors, he makes up for in macho.
\end{quote}
\doublespacing

In this study, we extend our prior work in Connolly and Rendleman (2008, henceforth CR) to address the issue of Tiger Woods' dominance on the PGA Tour, his effect on the play of other tournament participants, including those with whom he is paired, and Phil Mickelson's alleged propensity to play poorly in golf's major championships.  In CR, we use a generalized additive model to estimate time-dependent mean skill functions and the first-order autocorrelation of residual scores about their means for 253 active PGA Tour golfers over the 1998-2001 period.  In estimating these functions, we remove the estimated (random) effects associated with the relative difficulty of the course on which each round is played and the relative advantage or disadvantage of each player in playing these courses.  Although the CR data is somewhat dated, it covers a period of time when Woods had become known as an intimidating force in golf and Mickelson was being accused of choking in majors.

As expected, we find that Woods' play dominated that of other top PGA Tour professionals during our sample period.  During year 2000, generally acknowledged as Woods' best on the Tour, Tiger could have won a few tournaments with no luck at all and many more with just a little bit of luck.  By contrast, even his strongest competitors, including David Duval and Phil Mickelson, could not have come close to winning tournaments without a substantial amount of luck. Also, as expected, we find statistically significant adverse effects associated with being paired with Woods in the final round of a PGA Tour event.  Contrary to popular belief, however, those who are paired with Woods perform better when both are in contention to win a tournament than when they are not.

Interestingly, we find that Mickelson actually played well in golf's majors during the 1998-2001 period and that his level of play, relative to his norm, was comparable to that of Tiger Woods, who won five of 16 majors during this period, including the ``Tiger Slam,'' where he won four majors in a row -- the U.S. Open, British Open and PGA Championship in 2000 followed by the Masters in 2001.  Moreover, although Mickelson never won a major from 1998 through 2001, we show that he played with a sufficient degree of luck, or his temporary abnormal performance was sufficiently favorable, to have won as many as two major golf championships during this period.  In other words, bad luck, or more accurately, the good luck of others, played a substantial role in Mickelson's failure to win any of golf's majors between 1998 and 2001.

\subsection{PGA Tour Tournament Play and Qualifying for the Tour}

With a few exceptions, in a typical PGA Tour stroke-play event, 156 players begin play in the first of four 18-hole rounds of competition.  After the first two rounds, the 70 players and ties with the lowest 36-hole total scores ``make the cut'' and qualify for play in the last two rounds.  After the final round of play, the player with the lowest total 72-hole score wins the tournament.  If two or more players are tied for first place after 72 holes, the winner is determined by a playoff.  All players who make the cut and complete all four rounds of play earn prize money.  Those who miss the cut win no money.

At the end of each PGA Tour season, the 125 players who have won the most official prize money become ``fully exempt'' to participate on the Tour the following season.  Those ranked in positions 126-150 in money winnings become ``partially exempt,'' and are eligible to participate the next season on a limited basis.  All tournament winners in a given season earn fully-exempt status the following two years.  Those who win The Tour Championship and any World Golf Championship events earn fully-exempt status for three more years, and those who win the four majors and The Player's Championship earn fully-exempt status for five additional years.  In addition, approximately 50 players each year earn exemptions though the PGA Tour's annual qualifying tournament (``Q-school'') and by finishing among the top 25 money winners on the PGA Tour-sponsored Nationwide Tour.

\section{Data}
We collected individual 18-hole scores for every player in every stroke play tournament on the PGA Tour for years 1998-2001 for a total of 76,456 scores distributed among 1,405 players.  After limiting our sample to players who recorded more than 90 scores, the resulting sample employed here and in CR consists of 64,364 observations of 18-hole golf scores for 253 active PGA Tour players over 181 stroke-play events.  As we describe in CR, most of these omitted players are not representative of typical PGA Tour players.  For example, 565 of the omitted players recorded only two 18-hole scores. (More detailed characteristics of the sample are provided in our other paper.) By excluding these players, we maximize the power of Wang's (1998) cubic spline fitting methodology and minimize potential distortions in estimating the statistical properties of golf scores of regular players on the Tour.

\section{Statistical Estimation Model}
Our model for estimating the skill and luck components of PGA Tour players' golf scores is described in detail in CR.  Based on Wang (1998), we estimate a cubic spline-based time-varying mean skill level for each player, after reducing each player's 18-hole score by estimated random round-course effects common to all players and estimated random player-course effects.  These estimated effects capture the tendency of individual players to perform better or worse than normal on specific courses.  In simplified form, the model can be expressed as follows:
\begin{equation}
\
s_i  = h_i \left(  \bullet  \right) + r + c_i  + \theta _i
\end{equation}

In (1), $s_i$ is a given 18-hole score for player $i$.  $h_i \left(  \bullet  \right)$ is the cubic spline-based estimate of the same score after removing $r$, the random effect that captures the relative difficulty of the course on which the round was played on that particular day and  $c_i$, the random player-course effect that reflects the extent to which the course played favorably or unfavorably for player $i$.  $\theta _i$ is the residual random error term that is assumed to follow an AR(1) process, which is captured in the estimation of the cubic spline function $h_i \left(  \bullet  \right)$.  This residual error can be decomposed into two components, $\theta _i  = \lambda _i  + \eta _i$, where $\lambda _i$ represents the autocorrelated component and $\eta _i$ is white noise.  The model is estimated simultaneously for all 253 players in our sample.

Note that a round-course interaction is defined as the interaction between a regular 18-hole round of play in a specific tournament and the course on which the round is played.  For 156 of 181 tournaments in our sample, only one course is used and, therefore, there is only one such interaction per round.  By contrast, the first three rounds of the ATT Pebble Beach National Pro Am are played on three different courses using a rotation that assigns each tournament participant to each of the three courses over the first three days of competition.   A cut is made after the third round, and a final round is played the fourth day on a single course.  Thus, the Pebble Beach tournament consists of 10 round-course interactions - three for each of the first three days of competition and one additional interaction for the fourth and final day.

It should be noted that we do not include specific information about playing conditions (e.g., adverse weather as in Brown (2008), pin placements, morning or afternoon starting times, etc.) when estimating (1).  Nevertheless, if such conditions combine to produce abnormal scores in a given 18-hole round, the effects of these conditions should be reflected in the estimated round-course-related random effects.

Using (1), the estimated random effects associated with round-course interactions range from $-3.92$ to 6.95 strokes per round, implying almost an 11-stroke difference between the relative difficulty of the most difficult and easiest rounds played on the Tour during the 1998-2001 period.  By contrast, the estimated random effects associated with player-course interactions are very small, ranging from $-0.065$ to 0.044 strokes per round, an insufficient amount to have any impact on the overall scores in a typical 72-hole PGA Tour event.

	Much of the focus in this paper is on the estimated spline fits for individual players and associated $\theta$ and $\eta$ errors.  When we refer to a player playing ``normal,'' we are referring to a situation in which his 18-hole score for a given round, adjusted for estimated round-course and player-course effects, equals the time-dependent mean score for that same round given by the player's estimated cubic spline function.  Thus, a player who plays ``normal'' plays exactly as predicted by his estimated time-dependent mean skill level rather than at a level that might be characteristic of PGA Tour players in general.  $\theta$ errors can be viewed as differences between actual 18-hole scores and predicted scores, taking into account time-dependent cubic spline-based estimates of mean skill levels and the random effects associated with round-course and player-course interactions but without adjusting for autocorrelation in random error structures.  $\eta$ errors represent $\theta$ errors adjusted for first-order autocorrelation.

Although the potential for autocorrelated $\theta$ errors comes into play in estimating each player's cubic spline function, we do not focus on the correlation in residual errors in this paper.  However, in CR, we show that the first-order autocorrelation coefficient in $\theta$ errors is positive for 155 of the 253 players in our sample.  Only two players show evidence of significant negative autocorrelation in $\theta$ residuals, and 20 to 23 players (depending on the test) show evidence of significant positive autocorrelation.  Using the methodology of Ljung and Box (1978), we determine that there is no significant higher-order autocorrelation in $\eta$ residuals -- once we adjust for first-order autocorrelation in $\theta$ residuals, the remaining residual error for all 253 players represents white noise.

Spline fits for Tiger Woods, David Duval and Phil Mickelson, three golfers whose performance is analyzed in this study, are shown in Figure 1.  Actual 18-hole scores, reduced by random round-course and player-course effects, are plotted about the smooth spline fits.  The spline fit for Tiger Woods is U-shaped, reaching its minimum in year 2000, regarded as the best year in Woods' professional career.  Spline fits for David Duval and Phil Mickelson turn out to be linear, with Duval's being positively sloped, indicating his skill was deteriorating, and Mickelson's being negatively sloped, indicating an improvement in skill.  (Over all 253 players, 105 spline fits are exactly linear.) ``Scaled golf time'' for each player represents the chronological sequence of rounds for the player scaled to the \{0, 1\} interval.  Although all three players participated on the PGA Tour during roughly the same period of time, the scaled golf times for each player represent the joint sequencing of their own scores rather than the sequencing of the scores of the three players.

Visual inspection if Figure 1 reveals a very low signal-to-noise ratio for each individual player's spline fit relative to his 18-hole scores adjusted for estimated random round-course and player-course effects.  Although the pseudo adjusted R-square for the model as a whole is 0.296, calculated as $1 $-$ Mean\;square\;error / Mean\;square\;total$, the pseudo adjusted R-squares for the individual spline fits are not nearly as high.   For example, for Woods, Duval and Mickelson, the pseudo adjusted R-squares associated with their respective spline fits are 0.070, $-0.003$ and 0.034, with the highest being 0.408 for Billy Ray Brown.

Based on the relatively low individual pseudo adjusted R-squares, one might argue for a model simpler than a cubic spline function for estimating player skill.  Despite the low signal-to-noise ratio, using the bootstrap we show in CR that the spline model is significantly superior (at the 5\% level) to a player's mean score, adjusted for round-course and player-course interactions, for 71 of 253 players.  It is significantly superior to a linear time trend for 25 of the players and to a quadratic time trend for 10 players.  In addition, the spline model is superior to a time-dependent mean-adjusted score that varies by calendar year for 13 players.  Even though a simpler functional form might adequately capture time variation in mean player skill for the great majority of players, the spline model should (approximately) capture that simpler functional form when appropriate, but also capture more complex forms of time-varying skill when such patterns arise in the data.

Figure 1 shows the greatest dispersion of 18-hole scores (reduced by random round-course and player-course effects) about their respective spline fits for Phil Mickelson, followed by David Duval and Tiger Woods.  In fact, the standard deviations of $\theta$ residual scores for these three players are 3.02, 2.82 and 2.46 strokes, respectively, with the range for all 253 players in our sample of 2.14 to 3.44 strokes per round.  Mickelson's standard deviation of 3.02 strokes ranks 15th highest, while Woods' standard deviation of 2.46 strokes ranks 22nd lowest.  Although Figure 1 shows that Duval and Mickelson were capable of shooting scores as low as Woods, they could not do so as consistently.  Moreover, Woods' worst scores were not generally as high as those of Duval and Mickelson.

\section{The Dominance of Tiger Woods}
	In our original study, we demonstrate that the average total $\theta$ residual winning score per tournament over the 181 tournaments in our sample was $-9.64$ strokes, with the total $\theta$ residual ranging from +0.13 strokes for Tiger Woods in the 1999 Walt Disney World Resort Classic to $-21.59$ strokes for Mark Calcavecchia in the 2001 Phoenix Open.   The 1999 Walt Disney Resort Classic is the only tournament in our sample won by a player with a positive total $\theta$ residual score, meaning that the winner played slightly worse than normal as estimated by our model.  We also demonstrate that most players among the 20 to 30 best finishers in a tournament experienced negative total $\theta$ residual scores.  If the $\theta$ residual is viewed as a ``luck factor,'' our results indicate that to win a tournament on the PGA Tour, one must experience a sufficient amount of good luck to not only shoot a low total score, but to also overcome the collective good luck of the other participants in the tournament.  Only Tiger Woods was sufficiently skilled to have won a PGA Tour event during our sample period by playing ``normal.''

	An interesting way of assessing Woods' dominance is to ask how well he would have placed in the tournaments in which he participated by simply playing ``normal,'' and, therefore, experiencing no good luck as estimated by our model.\footnote{This analysis does not take into account the possible effect that Wood's normal play might have on the play of others, such as that documented in Section 5.} During the 1999-2001 period, Woods participated in 85 regular stroke-play events on the PGA Tour and never missed a cut in any of these tournaments. By contrast, David Duval, generally regarded as the second-best player during this period, missed six cuts, and Phil Mickelson missed 15.  (Mickelson's missing nine more cuts than Duval most likely reflects the greater variability in Mickelson's scores (adjusted for random round-course and player-course effects) rather than his slightly higher spline-predicted score.)   Table 1 lists these 85 tournaments along with the winning score, Woods' score, his expected score and his total $\theta$ residual score.\footnote{Wood's actually participated in one more regular stroke-play event, the 1998 ATT Pebble Beach National Pro Am.  The first two rounds of this tournament were played in January, 1998, but the weather was so bad that the third and fourth rounds could not be completed.  A third and final round was postponed and re-scheduled for July 1998.  Many players, including Woods, chose not to play in this final round.  Because of the unusual nature of this event, and the fact that Woods chose not to complete the tournament, we have not included it in our analysis in Table 1. }   For example, in the 1998 Mercedes Championships, the first tournament listed in Table 1, the winning score was 271 (column 1) and Woods' score was 272 (column 2).  Woods' total residual score was  $-3.87$ strokes (column 4), so Woods played 3.87 strokes better than predicted.  If we subtract the $-3.87$ stroke total residual from Woods' total score, we obtain a total expected score of $272 - (-3.87) = 275.87$ strokes (column 3).  Thus, if Woods had played to his norm over the four rounds of the Mercedes Championships, he would have shot 275.87.  (We ignore the fact that golf is scored in integers and cannot involve fractional amounts.)  Columns 5 and 6 indicate that Woods finished in a two-way tie for second in the tournament.  Column 7, the key column in this analysis, indicates that if Woods' total score had been 275.87 as predicted, he would have placed sixth overall.

	The average of values in column 7 indicate that if Woods had played as predicted in 1998, his average finish in the 19 stroke-play PGA Tour events in which he participated would have been 10.94, or 11th place.  In 1998, his best finish playing average would have been fourth.  A slightly improved pattern emerges in 1999, where he would have finished in sixth place on average by playing ``normal,'' and he could have won one tournament, which he did, by playing ``normal'' and finished among the top three in six more.

	There is one striking outlier among the 1999 PGA Tour events in which Woods participated - the ATT Pebble Beach National Pro Am - in which he would have placed 37th by playing ``normal.''  Otherwise, his worst finish in 1999 would have been seventh place.  The reason for this outlier is that the course rotation to which Woods was assigned played 5.44 strokes more difficult than the easiest rotation.  As shown in Table 2, only one player among the top 14 finishers in the tournament played a difficult course rotation (as indicated by a non-negative total round-course effect).  They were all blessed with a rotation that played approximately 5 strokes less difficult than Woods'.  By contrast, 10 players among the bottom 16 played one of the difficult rotation assignments as indicated by a total round-course effect greater than three strokes.  Note that no one among the top 14 finished the tournament with a positive total residual score, and no one among the bottom 16 finished with a negative total residual.  This is consistent with CR, where we show that almost all players who end up among the leaders in a PGA Tour event finish with negative total residual scores.

	Year 2000, generally regarded as Woods' best on the PGA Tour, is the year in which Woods won the final three majors of the year en route to his eventual ``Tiger Slam.''  Table 1 indicates that Woods could have won three tournaments in 2000 by playing at his (time-dependent) norm and that his average finish would have been between second and third place if he had played ``normal.''  Although Woods began year 2001 playing well, including winning the 2001 Masters to complete his ``Tiger Slam,'' his performance deteriorated in the second half of 2001 after making well-publicized changes to his swing.  On average, he would have finished between sixth and seventh place in 2001 by playing ``normal'' but would have done much better in the first half of the year than the second.

	We now contrast this performance to that of David Duval.  Duval was ranked third in the Official World Golf Rankings at the end of 1998, 2000 and 2001 and was ranked second at the end of 1999.  No other golfer was consistently ranked second when Duval was ranked third.  Moreover, if we rank players on the basis of their average spline-predicted score over our 1998-2001 sample period, Duval ranks second to Woods at 69.15 strokes vs. Woods' 68.18.

	Over our sample period, Duval missed six cuts; Woods missed none.   Although we provide no separate table for Duval, in the tournaments for which Duval made the cut, his average finish playing ``normal'' would have been position 9.32, 13.22, 13.25 and 17.53 in years 1999 to 2001, respectively.  Similarly, Phil Mickelson's average finish would have been position 30.81, 24.56, 14.50 and 9.95 by playing ``normal'' in the tournaments in which he made the cut.  (He actually missed 15 cuts during this period).  These results are in stark contrast to those of Woods, who could have won a handful of tournaments by playing ``normal'' and many more with just a little bit of luck.
\section{Psychological Pressure of Competing Against Tiger Woods}

	Over the years, we have heard many radio and television commentators claim that those who are paired with Tiger Woods tend to perform poorly, presumably due to the psychological pressure that Woods places on his fellow competitors.  Moreover, it is generally acknowledged that those who are paired with Woods in the final group on the final day of a PGA Tour event tend to do poorly relative to expectations (i.e., some would say that these players choke).\footnote{In PGA Tour events, the top two players at the end of the next-to-last day of play, generally the third round, are paired together in the last group in the fourth and final round.}

	In a recent paper, Brown (2008) posits that when competitors in tournaments or other settings such as the general workforce must compete against ``superstars,'' their productivity may be diminished because they feel there is no chance that they can compete successfully against the superstars.  She tests her hypothesis by examining the differential performance of PGA Tour players in tournaments in which Woods, the superstar, competes and does not compete over the period 1999 to 2006.  She concludes that there is a significant diminution of performance among regular PGA Tour players when Woods is also participating in the tournament.

	Tables 3 and 4 summarize all of our tests involving Woods' impact on the play of others in the field, including tests that address Brown's hypothesis directly.  In all tests, we regress $\eta$ residuals against one or more dummy variables involving the interaction of tournament participants with Woods.  We regress  $\eta$ residuals rather than $\theta$ residuals against dummies to more accurately isolate the pure impact of players' interactions with Tiger.  Otherwise, a portion of a player's abnormal performance in a given round would be attributable to the carryover of abnormal performance in previous rounds and could contaminate the effects we are trying to estimate.

	In our first test, summarized in Table 3, we regress $\eta$ residuals against a dummy variable indicating whether Woods is participating in the tournament for which the $\eta$ residual is estimated.\footnote{As reported in CR, the original model took 40 hours to estimate on a Windows XP-based PC with 1 GB RAM using a 2.80-GHz Intel Xenon processor.   Therefore, we do not re-estimate the CR model using the additional dummy variables in test 1 nor in any subsequent tests.  Moreover, we do not test for the significance of the dummy variable coefficients in the context of a re-estimated model, which would require the use of the bootstrap, the method of significance testing used in CR.  In CR, it took over five days to produce 40 bootstrap samples and, therefore, it would be impractical to employ a re-estimated model with accompanying bootstrap tests for the various estimation specifications employed in this study.}  This is a direct test of Brown's hypothesis, but we use a different dataset and an entirely different statistical methodology for estimating abnormal performance.  The coefficient on the dummy variable that indicates whether Woods is in the field is 0.051 strokes per round and has a p-value of 0.0211.  Thus, ignoring other factors involving Woods' interaction with other participants during tournament play, we find a statistically significant effect associated with having Woods in the field.  In general, scores of players other than Woods are 0.051 strokes higher (worse) in tournaments in which Woods participates.  Although statistically significant, this finding has little practical significance, since over the four rounds of play in a typical PGA Tour event, the total effect of having Woods in the field would be $0.051 \times 4 = 0.24$ strokes, an insufficient amount to change the final total score of any player whose score is 0.051 strokes per round higher due to Woods' presence in the tournament.  By contrast, Brown finds that when Woods is in the field, the performance of regular PGA Tour players deteriorates by 0.2 strokes per round.

In our second test, we separate players into two groups for tournaments in which Woods is in the field.  The first group includes those who are playing with Woods, and the second group includes those with whom Woods is not playing.  In these regressions of  $\eta$ residuals, the coefficient associated with playing with Woods is 0.478, and with a p-value of 0.0004, is statistically significant.  The coefficient associated with not being paired with Woods when he is in the field is 0.043, with a p-value of 0.0505.  Thus, it appears that Woods' adverse impact on the play of others is real and statistically significant for those with whom he is paired but very small for those with whom is is not paired.

	In test 3, we regress $\eta$ residuals against a dummy variable indicating whether the player associated with the $\eta$ residual score is actually playing with Woods in the same group.  This is a slightly different test than test 2, since all other players are treated the same, even if Tiger is not playing in the tournament.  The estimated coefficient is 0.462 with a p-value of 0.0005.  Consistent with test 2, being paired with Woods, rather than just having Tiger in the field, appears to have an adverse impact on player scores.

	In test 4, we examine the effects of playing with Woods on a round-by-round basis.  As is evident from the total winning scores shown in Table 1, all but two of the tournaments in which Woods participated in the 1998-2001 period involved four rounds of play, with the exceptions being the 1998 Buick Invitational and the 1999 ATT Pebble Beach National Pro Am, which were cut short due to adverse weather conditions.

Test 4 shows positive coefficients for each round in which players are paired with Woods, but the coefficients are statistically significant only for the first and fourth rounds, where the estimated coefficients are 0.611 and 0.857, respectively.  It should be noted that in almost all PGA Tour events, players who are paired together in the first round are also paired together in the second.  If we were forced to tell a story based on the results of test 4, the coefficient estimates suggest that players paired with Woods in the first round of a tournament may be nervous and intimidated, but by the second day they settle down and play more to their norm.  The coefficient of 0.857 for round 4 suggests that players succumb to more pressure playing with Woods on the final day of a tournament when the ``money is on the line.''

In test 5 we address the question of whether the apparent adverse effect of playing with Woods in round 4 is simply the effect of playing in the final round rather than the effect of playing with Woods.  In this test we separate scores from the final scheduled round of a tournament into two groups.\footnote{In all but the Bob Hope Chrysler Classic and the Las Vegas Invitational, which are 5-round tournaments, the final scheduled round is round 4.}  The first group includes scores of players who are playing in the final scheduled round with Woods, and the second group includes those playing in the final scheduled round but not with Woods.  The coefficient associated with playing in the final round without Woods is not significantly different from zero, while that associated with playing with Woods in the final scheduled round is 0.858 with a p-value of 0.0030.  Thus, there does appear to be an adverse effect associated with playing with Woods in the final round separate from any overall final round effect.

	To further test the hypothesis that  players may succumb to more pressure playing with Woods on the final day of a tournament when the ``money is on the line,'' we separate players paired with Woods in round 4 into two groups.  The first are players who are paired with Woods when he is ``in contention,'' with ``in contention'' defined in various ways, below.  The second are players paired with Woods when he is not in contention.

Our first definition of ``in contention'' is Woods playing in the final group in a final scheduled tournament round.  In all PGA Tour stroke-play events, after the cut, the order of play in subsequent rounds is determined by a player's position at the end of the previous round.  After the cut, those who lead a tournament at the end of round $t$ are paired in the final group and tee off last on day $t + 1$.  Thus, if a player is playing in the last group with Woods on the final day of a tournament, the two must be among the top two players in the field entering the final round and, obviously, with the exception of others who may be tied with Woods and his playing partner(s), are in the best position to win the tournament.\footnote{In some tournaments, where there is a risk that play will not be completed before dark, three players per group may compete in the final round.}

	One might think that if there was ever a time of intimidation associated with playing with Woods, it would be when one is paired with Woods on the final day in a strong position to win a tournament.  However, the data don't bear this out.  The initial specification of test 6 shows that those who play with Woods on the final day of a tournament when Woods is \emph{not} in the final group shoot 1.122 strokes worse than normal (p-value = 0.0019).  By contrast, those who are paired with Woods in the final round when he \emph{is} playing in the final group shoot only 0.389 strokes worse than normal, which, with a p-value of 0.4184, is not statistically different from zero.  We note that there are only 31 instances of players being paired with Woods in the final group during the 1998-2001 period, and, therefore, we may not have a powerful test.

We also define Woods being in contention as Tiger being within ten, eight, six, four, two or zero strokes of the lead going into the final round.  As shown in the second through fifth specifications of test 6, if being in contention is defined as Woods being within four to ten strokes of the lead, those who are paired with Tiger in the final round when he is in contention score 0.80 to 0.88 strokes worse than normal, with p-values ranging from 0.0087 to 0.0297.  Note that this effect is essentially the same as the 0.857 adverse strokes per round associated with simply playing with Tiger in the final round (test 4).  In essence, since Woods is almost always within four to ten strokes of the lead going into the final round of a tournament, defining Woods ``being in contention'' in this fashion is hardly more than identifying that Tiger is playing in the final round.

The final two specifications of test 6 show that when ``being in contention'' is defined as Woods being within two strokes of the lead or tied for the lead or better, the adverse affect associated with playing with Tiger in the final round falls to 0.690 and 0.329 strokes, respectively, but neither estimate is statistically significant.  When these results are paired alongside those where ``being in contention'' is defined as Woods playing in the final group, contrary to conventional wisdom, we find no statistically significant adverse affect associated with being paired with Tiger in the final round at times when Tiger is truly in a position to win.  Perhaps the best explantation for this misperception is that Woods is so much more skilled than those with whom he might be paired in the final round of a tournament that when his playing partners play close to their norm, compared to Woods they appear to be playing poorly.\footnote{It is important to note that in a typical PGA Tour event, player pairings for the first two rounds are determined prior to the start of play and do not reflect tournament performance.  However, after the cut (typically after round 2), player pairings are based on cumulative performance in the tournament's previous rounds.  Generally, those who are paired together have recorded the same score, or very close to the same score, going into the round for which they are paired.  Since Tiger Woods is so highly skilled, if a player is paired with Woods in the third or fourth rounds, he probably played with an exceptionally degree of good luck during the tournament's previous rounds.  Thus, on average, those who are paired with Woods in the third or fourth rounds of a tournament should have had a negative cumulative tournament residual score prior to being paired with Woods.  Since $\theta$ residual scores must sum to zero, and $\eta$ residuals sum close to zero (on average, over all 64,364 observations, the average $\eta$ is 0.00006), an expected negative cumulative tournament $\eta$ residual, conditional on being paired with Woods in the third or fourth round, implies a positive expected residual in subsequent rounds.  This is not because players who are paired with Woods tend to choke, but because $\eta$ residuals must sum (close) to zero. In a second set of tests whose results we do not report, we adjust residual scores for rounds in which players are paired with Woods by an amount that reflects the player's performance in the previous rounds of the same tournament. For example, consider a player who records a total of 102 scores in our four-year sample period.  Assume this player is paired with Woods in round 3 of a particular tournament. During rounds 1 and 2 of the same tournament, the player's $\eta$ residual scores are $-2$ and $-3$, respectively.  Since the sum of the player's $\eta$ residual scores must be close to zero, we would expect the remaining 100 $\eta$ residual scores for the same player to sum to 5, or 0.05 strokes per round.  Therefore, in this instance, we would subtract 0.05 strokes from the player's actual third-round $\eta$ residual when computing the effect of being paired with Woods in the third round, and we would make similar adjustments to $\eta$ residual when running regressions involving pairings with Woods in the fourth round. As it turns out, these adjustments make little if any difference in our test results.  Consistent with our assumption that players who are paired with Woods should have recorded negative $\eta$ residual scores on average in a tournament's previous rounds, all coefficients associated with being paired with Woods are lower than those reported in Table 3.  However, at most, the coefficients are reduced by 0.03 strokes, and all that are statistically significant prior to the adjustment remain so after the adjustment.}

It should be noted that when a player is paired with Woods in the final round of a tournament and Woods is within `X' strokes of the lead, the player paired with Woods is probably also within `X' strokes of the lead or very close.  Therefore, it is possible that any adverse effect on a player's score when paired with Woods in the final round has nothing to do with Woods but, rather, reflects the effect of being within `X' strokes of the lead.  To test this hypothesis, we ran a series of six regressions of $\eta$ residuals in final scheduled tournament rounds against dummy variables reflecting whether a player (not Woods) is or is not in contention and whether he is paired with Woods.\footnote{Some tournaments are cut short due to adverse weather conditions.  In most cases, a decision is made to end the tournament after the next-to-last scheduled round is played.  Since our intent is to estimate the effect of being in contention, if a tournament is cut short, a player who turned out to be in contention going into the final round most likely did not know it at the time.  Therefore, in these tests, we only consider final round scoring in tournaments played the full number of originally scheduled rounds.}  The six regressions vary according to the specification of being in contention, with ``in contention'' defined as being within ten, eight, six, four, two or zero strokes of the lead in tests 7-12, respectively.  Coefficient estimates and associated p-values are summarized in Table 4.

In all tests, the coefficient associated with being paired with Tiger Woods is statistically significant and of the same order of magnitude as in test 5 (summarized in Table 3). In tests 8-10, where being in contention is defined as being within four to eight strokes of the lead, the ``in contention'' coefficient is also significant and on an order of magnitude of 0.104 to 0.171 strokes per round.   In tests 5 and 6, where ``in contention'' is defined as being within two stokes of the lead or actually being in the lead going into the final scheduled round, the estimated ``in contention'' coefficients are slightly higher but insignificant, most likley due to insufficient observations.  Overall, the results suggest that those who are in contention going into the final scheduled round of a tournament with some possibility of winning play a little worse than normal.  This might be due to nervousness or a tendency to take more risks, which could lead to higher scores.

Overall, our results contrast with those of Guryan, Kroft and Notowidigdo (2008) who study the effect of pairings in PGA Tour events during the 2002, 2005 and 2006 seasons.  Examining all parings, not just those involving Tiger Woods,  these authors find no evidence that the ability of playing partners affects the performance of professional golfers.  For pairings involving Woods, they find that players perform 0.354 strokes per round \emph{better} than average, but this amount is not statistically significant.  Guryan, Kroft and Notowidigdo use a statistical method for estimating abnormal performance entirely different from ours and estimate the effects of player pairings over a different period of time.  Thus, our findings and theirs may not be directly comparable.

\section{How Poorly did Phil Mickelson Perform in ``Majors''}

	Until Phil Mickelson won the 2004 Masters, he was accused, almost unmercifully, of choking in the four major golf championships.\footnote{A Google search on the strings "mickelson" and "choke" yields 23,000 references.}  Despite his stellar record in other PGA Tour events, his lack of success in majors would have defined him as a good player, but not a great player, if he had never won a major championship in golf.  At the same time, it was generally understood that Tiger Woods stepped up his game in major championships, winning a total of five out of a possible 16 over the 1998-2001 period.

	Table 5 summarizes Phil Mickelson's performance in the 16 major golf championships over our four-year sample period.  Mickelson missed the cut in the 1999 British Open but, otherwise, made the cut in the remaining 15 events.

	Column 1 shows the winning score for each of the 16 tournaments, and column 2 shows Mickelson's score. Phil's total four-round $\theta$ residual score is shown in column 4.  (Here, we focus on $\theta$ residuals, rather than $\eta$ residuals, because we are concerned about the extent to which Mickelson played better or worse than normal, regardless of whether his abnormal performance resulted from a carryover of abnormal performance from previous rounds.)  Subtracting this total residual from his actual score gives his expected score in column 3.  For example, in the 1998 Masters, Mickelson's actual total score was 286, and his $\theta$ residual score was $-4.66$.  Thus, Mickelson played a total of 4.66 strokes better than normal over the four rounds of play.  If he had played ``normal'' his four-round total would have been $286 - \left( { - 4.66} \right) = 290.66$, the value shown as his expected score in column 3.

	Note that in two of the 16 tournaments, Mickelson's total residual score was very low, $-12.31$ strokes in the 1999 U.S. Open and $-9.49$ strokes in the 2001 PGA Championship.  As we demonstrate in CR, the average total $\theta$ residual winning score per tournament over the 181 tournaments in our sample was $-9.64$ strokes.  Thus, Mickelson, being more highly skilled than the typical winner of a PGA Tour event, could have won most tournaments in which he played if his total four-round residual score had been $-9.49$ strokes or better, assuming, of course, that the scores of all other tournament participants would have remained the same..

	 We now address the question of how many majors Mickelson could have won by performing at the level associated with his $-12.31$ four-round total residual score in the 1999 U.S. Open.  For example, Mickelson's expected score in the 1998 Masters was 290.66.  If his residual score in the 1998 Masters had been $-12.31$ as in the 1999 U.S. Open, Mickelson's four-round total score would have been $290.66 - 12.31 = 278.35$.  This is lower than the actual winning score of 279, so Mickelson would have won the 1998 Masters by playing at the same level relative to his norm as he did in the 1999 U.S. Open (ignoring that golf is not played with fractional scores and that his higher level of play would have had no impact on the play of others in the tournament).  Stated differently, the degree of luck, or temporary abnormal performance, that Mickelson experienced in the 1999 U.S. Open would have been sufficient to have enabled him to win the 1998 Masters.  Applying this same logic to all 15 majors for which Mickelson made the cut, we see that Phil's abnormal performance in the 1999 U.S. Open, if applied to the other majors, would have enabled him to win 11 of 16 times.  Unfortunately for Phil, Payne Stewart's total $\theta$ residual score of $-14.301$ strokes in the 1999 U.S. Open (not shown in Table 5), was a sufficient departure from his norm to have enabled Stewart to place one stroke ahead of Phil.  Did Mickelson choke? Certainly not, but, apparently, he did not have quite as much good luck as Payne Stewart.

	If we apply Mickelson's total residual score of $-9.49$ strokes in the 2001 PGA Championship to all 15 tournaments in which he made the cut, we see that he played sufficiently well to have won three.  Although not as compelling as his performance in the 1999 U.S. Open, the results summarized in Table 5 suggest that Phil played well enough during the 1998 to 2001 period to have won one or two major golf championships.

	Table 6 shows the average $\theta$ residual score in the 16 majors for both Woods and Mickelson on a round-by-round basis.  Note that both tended to play better than their norms in the first two rounds.  Woods performed an average of 0.632 and 0.610 strokes better than normal in the first two rounds of majors compared with Mickelson's 0.903 and 1.421-stroke superior-than-normal performance.  Thus, in the first two rounds, Mickelson actually played better than Woods, relative to his norm.

	The story changes somewhat in the third and fourth rounds.  Woods tended to play closer to his norm in the final two rounds, and Mickelson played approximately one-half stroke worse than his norm.  Perhaps those who accused Mickelson of choking in the final two rounds of major championships did not realize that he had generally played exceptionally well in the first two rounds but had gotten back to (roughly) normal in the second two.  Playing one-half stroke per round worse than normal should not be considered choking.

	Overall, Woods played 0.382 strokes better than normal in the 16 majors, and Mickelson's performance averaged 0.317 strokes better than his norm.  To ensure that these averages are not dominated by outliers, we calculated the proportion of rounds in majors that both players played better than ``normal.'' Woods recorded a negative $\theta$ residual in 35 or 64 total rounds in majors, or 54.7\% of his rounds, while Mickelson played better than ``normal'' in 37 of 62 rounds, or 59.7\% overall.  Thus, the general consensus that Woods stepped up his game in majors and Mickelson choked is hardly fair, at least from 1998 to 2001.

Compared with Mickelson, Woods' success in majors lay mainly from his superior skill level.  Over our entire four-year sample period, the average value of Woods' spline-based estimated skill was 68.18 strokes per round.  The same average for the next-best-player, David Duval, was 69.15, almost a full stroke difference.  Mickelson's average was 69.51.  Therefore, Woods had an average $69.51 - 68.18 = 1.33$ per round stroke advantage over Mickelson based on skill alone, or $1.33 \times 4 = 5.32$ strokes over a full four-round tournament.  This is a lot of ground to have to make up in golf.  So it is not surprising that by playing in majors roughly at the same skill levels relative to their norms, Woods won five majors and Mickelson won none.  It is unfortunate, however, that Mickelson was characterized as an incomplete player who could not handle the pressure of golf's major championships.
\section{Summary and Conclusions}

Using the model in Connolly and Rendleman (2008), we demonstrate that by playing ``normal,'' Tiger Woods could have won some tournaments and placed no worse than fourth in the tournaments in which he participated in year 2000, his best on the PGA Tour.  More generally, his average finishing position would have been 11th, 6th, 2nd, and 7th in years 1998-2001, respectively, if he had played ``normal.'' No other PGA Tour player in our sample came close to such a feat.

We also quantify the intimidation factor associated with playing with Woods.  On average, players who were paired with Woods during the 1998-2001 period scored 0.462 strokes per round worse than normal.  Although we find that Wood's presence in a tournament may have had a small, but statisically significant adverse impact on the entire field, this effect was swamped by the apparent intimidation factor associated with having to play with Woods side-by-side.  However, contrary to popular belief, the adverse effect associated with being paired with Woods was the smallest when Woods and his playing partners were in contention to win.

It is also commonly held that Phil Mickelson performed poorly in majors prior to winning the Masters in 2004.  However, our data suggest the opposite.  Although Mickelson won no majors during our 1998-2001 sample period, he played sufficiently well to have won one or two majors under normal circumstances.  Moreover, his overall performance in majors, relative to his estimated skill level, was comparable to that of Tiger Woods, the winner of five major golf championships from 1998 to 2001.  Thus, the general characterization of Woods as golf's dominant player over the 1998-2001 period was accurate, but the frequent characterization of Phil Mickelson performing poorly in majors and choking was not.

We believe that the methods used in the analysis here should lend themselves favorably to modeling performance in a number of other sports.  The list of sports where athletes compete against one another indirectly is substantial: skiing, track and field, bowling, diving, equestrian, figure skating, gymnastics, rowing, shooting, swimming, weightlifting, and yachting.  In each case, the athletes do not have to contend with direct play of competitors as in basketball, volleyball, or tennis, but compete against a course and other athletes indirectly.  In some settings, the importance of random effects may be very small.  In diving, for example, the board is the same height for everyone.  Unlike golf, there is no variation in the athletic environment.  In skiing, however, athletes compete on multiple mountains over a season, and weather and course conditions may vary over the term of a competition.  This suggests that controls for these effects might be important.

Besides measuring the relative importance of skill and luck in athletic performance, the methods we used might also be used to construct athlete rankings.  Using a proper model, the spline fit at a point in time is a measure of athletic performance that accounts for multiple factors affecting measured outcomes.  We believe it would prove to be an interesting exercise to compare rankings generated by proper statistical models of athletic performance to those commonly used.

\clearpage
\pagestyle{empty}
\singlespacing
\begin{center}
\large{\textbf{References}}

\end{center}

\begin{small}

\noindent
Brown, J., 2008, ``Quitters Never Win:  The (Adverse) Incentive Effects of Competing with Superstars,'' working paper, Department of Agricultural and Resource Economics, University of California at Berkeley (April).\vspace{0.12in}

\noindent
Connolly, R. A. and R. J. Rendleman, Jr., 2008, ``Skill, Luck and Streaky Play on the PGA Tour,'' {\em The Journal of The American Statistical Association} 103, 74-88.\vspace{0.12in}

\noindent
Gladwell, B., 2008, ``How to Intimidate Golfers Like Tiger Woods,'' {\em Ezine Articles},\\ URL: http:\/\/ezinearticles.com\/?How-to-Intimidate-Golfers-Like-Tiger-Woods\&id=1167718 (May 10).\vspace{0.12in}
\noindent

\noindent
Golf Today, 2000, ``Tiger Woods Geared up for Masters Challenge,'' {\em golftoday.co.uk},\\ URL: http:\/\/www.golftoday.co.uk\/tours\/2000\/masters\/preview11.html.\vspace{0.12in}

\noindent
Guryan,J, K. Kroft and M. Notowidigdo, 2008, ``Peer Effects in the Workplace: Evidence from Random Groupings in Professional Golf Tournaments,'' NBER working paper.\vspace{0.12in}

\noindent
Ljung, G. M. and Box, G. E. P., 1978, ``On a Measure of Lack of Fit in Time Series Models,'' {\em Biometrika} 65, 297 - 303.\vspace{0.12in}

\noindent
Smith, M., 2001, ``Mickelson Seeks Elusive Major,'' {\em CBCsports.ca},\\ URL: http:\/\/www.cbc.ca\/sports\/story\/2001\/08\/15\/mickelson010815.html  (August 15).\vspace{0.12in}

\noindent
Wang, Y., 1998, ``Smoothing Spline Models with Correlated Random Errors,'' {\em The Journal of The American Statistical Association} 93, 341-348.\vspace{0.12in}

\end{small}
\clearpage


\begin{figure}[loc=h]
\centerline{\includegraphics[width=6in]{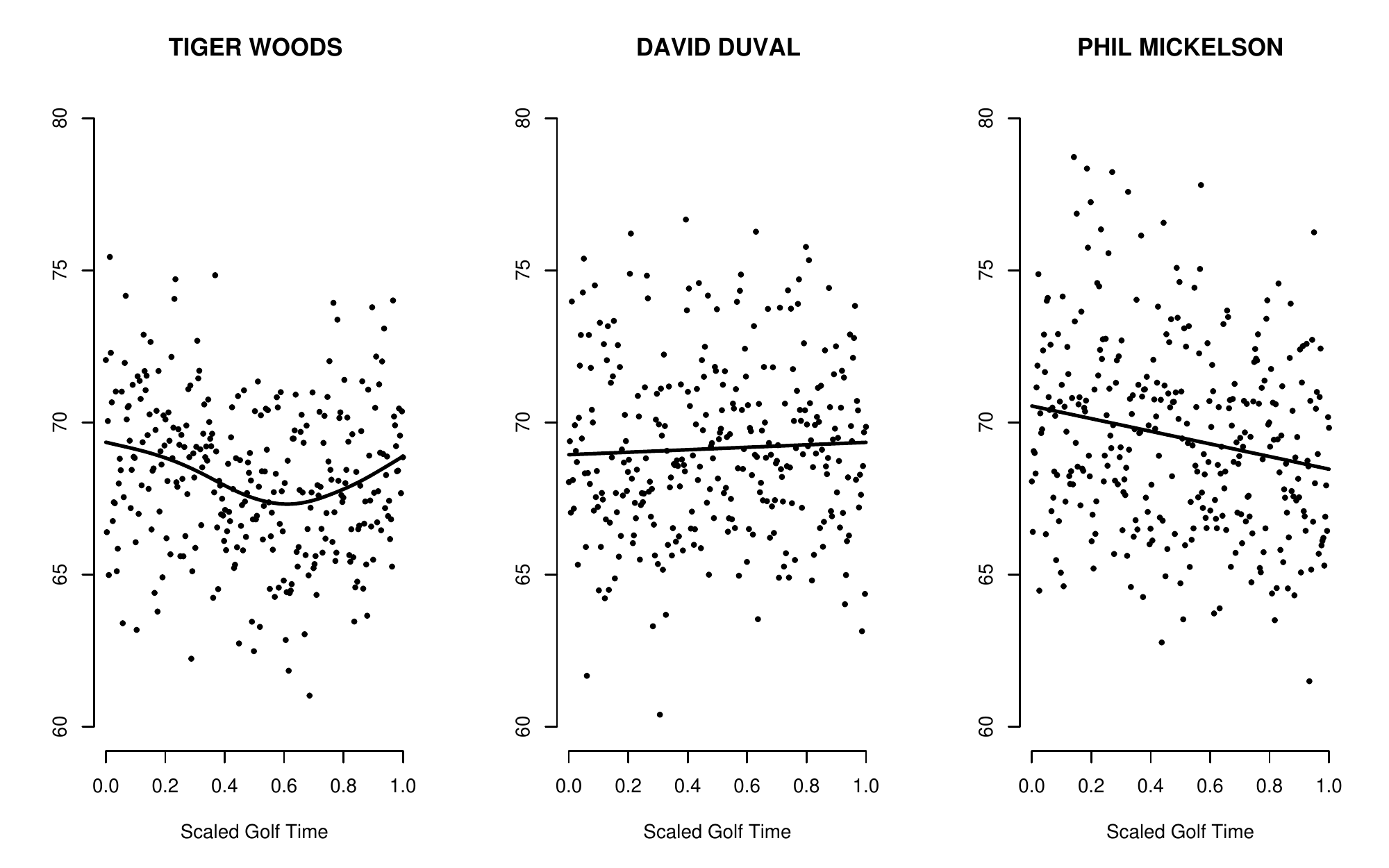}}

\caption{Spline-based estimates of mean skill levels.}
\end{figure}

\vspace{25mm}
Plots show 18-hole scores reduced by random round-course and player-course effects along with corresponding spline fits (smooth lines).  Scaled golf time for each player represents the chronological sequence of rounds for the player scaled to the \{0, 1\} interval.

\clearpage

\begin{table}[loc=h]
\caption{How Tiger Woods would have Placed in PGA Tour Stroke-play Events by Playing ``Normal''}
\label{id}
\begin{center}

\begin{tabular}{lrrrrrrr}
\hline
           &            &            &    Woods' &   Woods'  &            &    Players &    Woods' \\

           &    Winning &    Woods'  &   expected &   residual &    Woods'  &  tied with &      place \\

           &      score &      score &      score &      score &      place &      Woods  &  if ``normal'' \\

           &        (1) &        (2) &        (3) &        (4) &        (5) &        (6) &        (7) \\
\hline
98 Mercedes &        271 &        272 &     275.87 &      -3.87 &          2 &          2 &          6 \\

98 Buick Inv &        204 &        205 &     208.09 &      -3.09 &          3 &          3 &         21 \\

 98 Nissan &        272 &        272 &     279.75 &      -7.75 &          1 &          2 &         15 \\

  98 Doral &        278 &        283 &     283.71 &      -0.71 &          9 &          6 &         15 \\

98 Bay Hill &        274 &        284 &     283.77 &       0.23 &         13 &          4 &         13 \\

98 Players Champ &        278 &        290 &     286.17 &       3.83 &         35 &          7 &         18 \\

98 Masters &        279 &        285 &     285.86 &      -0.86 &          7 &          4 &         11 \\

98 BellSouth &        271 &        271 &     277.87 &      -6.87 &          1 &          1 &         14 \\

98 Byron Nelson &        265 &        272 &     271.26 &       0.74 &         12 &          3 &         12 \\

98 Memorial &        271 &        288 &     279.21 &       8.79 &         51 &          6 &         11 \\

98 U.S. Open &        280 &        290 &     286.85 &       3.15 &         17 &          5 &          7 \\

98 Western &        271 &        281 &     280.87 &       0.13 &          9 &          8 &          9 \\

98 British &        280 &        281 &     288.42 &      -7.42 &          3 &          1 &         10 \\

98 Buick Open &        271 &        275 &     274.14 &       0.86 &          4 &          2 &          4 \\

98 PGA Champ &        271 &        279 &     279.92 &      -0.92 &         10 &          3 &         13 \\

    98 NEC &        269 &        275 &     276.23 &      -1.23 &          5 &          2 &          9 \\

 98 Disney &        272 &        277 &     275.55 &       1.45 &          7 &          4 &          4 \\

98 Tour Champ &        274 &        289 &     279.14 &       9.86 &         20 &          1 &          5 \\

99 Mercedes &        266 &        277 &     277.38 &      -0.38 &          5 &          3 &          8 \\
 &            &            &            &            &            &            &            \\

           &            &            &            &            &            &    Average &      10.94 \\

           &            &            &            &            &            &            &            \\

99 Phoenix &        273 &        276 &     281.19 &      -5.19 &          3 &          1 &          6 \\

    99 ATT &        206 &        219 &     216.01 &       2.99 &         51 &         13 &         37 \\

99 Buick Inv &        266 &        266 &     272.57 &      -6.57 &          1 &          1 &          4 \\

 99 Nissan &        270 &        272 &     273.49 &      -1.49 &          2 &          3 &          7 \\

99 Bay Hill &        274 &        290 &     278.54 &      11.46 &         55 &          4 &          5 \\

99 Players Champ &        285 &        291 &     288.86 &       2.14 &         10 &          7 &          4 \\

99 Masters &        280 &        289 &     284.39 &       4.61 &         17 &          5 &          6 \\

    99 MCI &        274 &        280 &     274.36 &       5.64 &         17 &          9 &          4 \\

99 Byron Nelson &        262 &        271 &     267.58 &       3.42 &          7 &          2 &          3 \\

99 Memorial &        273 &        273 &     278.05 &      -5.05 &          1 &          1 &          3 \\

99 U.S. Open &        279 &        281 &     285.33 &      -4.33 &          3 &          2 &          6 \\

99 Western &        273 &        273 &     276.59 &      -3.59 &          1 &          1 &          3 \\

99 British &        290 &        294 &     291.49 &       2.51 &          5 &          3 &          4 \\

99 PGA Champ &        277 &        277 &     280.89 &      -3.89 &          1 &          1 &          5 \\

    99 NEC &        270 &        270 &     272.57 &      -2.57 &          1 &          1 &          3 \\

 99 Disney &        271 &        271 &     270.87 &       0.13 &          1 &          1 &          1 \\

99 Tour Champ &        269 &        269 &     269.39 &      -0.39 &          1 &          1 &          2 \\

99 Am Express &        278 &        278 &     279.32 &      -1.32 &          1 &          2 &          3 \\
 &            &            &            &            &            &            &            \\

           &            &            &            &            &            &    Average &          6.00 \\

\end{tabular}
\end{center}
\end{table}

\clearpage

\begin{center}

\begin{tabular}{lrrrrrrr}

\multicolumn{ 1}{r}{} &            &            &    Woods'  &    Woods' &            &    Players &    Woods'  \\

\multicolumn{ 1}{r}{} &    Winning &    Woods' &   expected &   residual &    Woods'  &  tied with &      place \\

\multicolumn{ 1}{r}{} &      score &      score &      score &      score &      place &      Woods  &  if ``normal'' \\

           &        (1) &        (2) &        (3) &        (4) &        (5) &        (6) &        (7) \\
\hline
00 Mercedes &        276 &        276 &     279.54 &      -3.54 &          1 &          2 &          3 \\

    00 ATT &        273 &        273 &     273.42 &      -0.42 &          1 &          1 &          2 \\

00 Buick Inv &        270 &        274 &     271.68 &       2.32 &          2 &          2 &          2 \\

 00 Nissan &        272 &        279 &     270.24 &       8.76 &         18 &          7 &          1 \\

00 Bay Hill &        270 &        270 &     274.29 &      -4.29 &          1 &          1 &          3 \\

00 Players Champ &        278 &        279 &     282.29 &      -3.29 &          2 &          1 &          3 \\

00 Masters &        278 &        284 &     280.77 &       3.23 &          5 &          1 &          2 \\

00 Byron Nelson &        269 &        270 &     269.33 &       0.67 &          4 &          2 &          4 \\

00 Memorial &        269 &        269 &     274.25 &      -5.25 &          1 &          1 &          4 \\

00 U.S. Open &        272 &        272 &     285.88 &     -13.88 &          1 &          1 &          2 \\

00 Western &        274 &        281 &     270.71 &      10.29 &         23 &          5 &          1 \\

00 British &        269 &        269 &     273.67 &      -4.67 &          1 &          1 &          2 \\

00 Buick Open &        268 &        275 &     267.46 &       7.54 &         11 &          5 &          1 \\

00 PGA Champ &        270 &        270 &     274.46 &      -4.46 &          1 &          2 &          3 \\

    00 NEC &        259 &        259 &     267.76 &      -8.76 &          1 &          1 &          2 \\

00 Canadian Open &        266 &        266 &     268.64 &      -2.64 &          1 &          1 &          3 \\

 00 Disney &        262 &        265 &     264.73 &       0.27 &          3 &          1 &          3 \\

00 Tour Champ &        267 &        269 &     269.69 &      -0.69 &          2 &          1 &          3 \\

00 Am Express &        277 &        281 &     277.05 &       3.95 &          5 &          3 &          2 \\

           &            &            &            &            &            &            &            \\

           &            &            &            &            &            &    Average &       2.42 \\

           &            &            &            &            &            &            &            \\

01 Mercedes &        274 &        280 &     273.21 &       6.79 &          8 &          4 &          1 \\

01 Phoenix &        256 &        271 &     269.21 &       1.79 &          5 &          2 &          4 \\

    01 ATT &        272 &        280 &     273.60 &       6.40 &         13 &          7 &          3 \\

01 Buick Inv &        269 &        271 &     269.28 &       1.72 &          4 &          1 &          4 \\

 01 Nissan &        276 &        279 &     272.35 &       6.65 &         11 &          7 &          1 \\

01 Bay Hill &        273 &        273 &     277.68 &      -4.68 &          1 &          1 &          3 \\

01 Players Champ &        274 &        274 &     282.62 &      -8.62 &          1 &          1 &         10 \\

01 Masters &        272 &        272 &     277.85 &      -5.85 &          1 &          1 &          4 \\

01 Byron Nelson &        263 &        266 &     264.84 &       1.16 &          3 &          3 &          3 \\

01 Memorial &        271 &        271 &     280.59 &      -9.59 &          1 &          1 &          6 \\

01 U.S. Open &        276 &        283 &     281.94 &       1.06 &         10 &          3 &          6 \\

01 Westchester &        268 &        280 &     275.79 &       4.21 &         16 &          3 &         10 \\

01 Western &        267 &        280 &     275.66 &       4.34 &         20 &         11 &          5 \\

01 British &        274 &        283 &     280.42 &       2.58 &         16 &          3 &         14 \\

01 PGA Champ &        265 &        279 &     275.74 &       3.26 &         26 &          7 &         12 \\

    01 NEC &        268 &        268 &     275.14 &      -7.14 &          1 &          2 &         10 \\

01 Canadian Open &        266 &        276 &     271.42 &       4.58 &         21 &          7 &          7 \\

 01 Disney &        266 &        272 &     270.61 &       1.39 &         14 &          5 &         11 \\

01 Tour Champ &        270 &        276 &     274.96 &       1.04 &         13 &          2 &         10 \\

           &            &            &            &            &            &            &            \\

           &            &            &            &            &            &    Average &       6.53 \\

\end{tabular}
\end{center}
\clearpage

\begin{table}[loc=h]
\caption{Actual Scores and Residual Scores for Selected Players in 1999 ATT Pebble Beach National Pro Am, Played Over a Total of Three Rounds}
\label{id}
\begin{center}

\begin{tabular}{lrrrrr}
\hline

           &            &        &     Total &    Total &            \\
          &            &        &     round- &    player- &            \\

           &            &      Total $\theta$ &     course &     course &      Total \\

    Player &      Score &   residual &     effect &     effect &   residual \\
\hline
\textbf{Top 14 }& \\
Payne Stewart &        206 &     -8.991 &     -1.859 &     -1.859 &    -12.708 \\

Frank Lickliter &        207 &    -10.796 &     -1.859 &      0.021 &    -12.634 \\

Craig Stadler &        209 &     -8.962 &     -1.336 &     -0.039 &    -10.336 \\

Fred Couples &        210 &     -5.585 &     -1.336 &     -0.021 &     -6.941 \\

Jay Williamson &        210 &    -10.808 &     -1.859 &     -0.035 &    -12.702 \\

Justin Leonard &        210 &     -4.531 &     -1.859 &     -0.007 &     -6.397 \\

Ronnie Black &        210 &    -10.296 &     -2.062 &     -0.023 &    -12.381 \\

Neal Lancaster &        211 &     -9.066 &     -1.336 &     -0.028 &    -10.430 \\

Tommy Tolles &        211 &    -10.380 &     -1.336 &     -0.009 &    -11.726 \\

Brett Quigley &        212 &    -13.248 &      3.195 &      0.010 &    -10.043 \\

Davis Love III &        212 &     -1.338 &     -1.336 &      0.001 &     -2.673 \\

Paul Azinger &        212 &     -5.119 &     -1.336 &     -0.018 &     -6.473 \\

Tim Herron &        212 &     -5.479 &     -1.336 &     -0.022 &     -6.836 \\

Vijay Singh &        212 &     -2.399 &     -1.859 &     -0.026 &     -4.284 \\

           &            &            &            &            &            \\

Tiger Woods &        219 &      2.995 &      3.382 &      0.008 &      6.385 \\

           &            &            &            &            &            \\
\textbf{Bottom 16} &  \\
 Guy Boros &        231 &      4.740 &     -2.062 &      0.005 &      2.683 \\

Omar Uresti &        231 &      5.882 &      3.382 &      0.011 &      9.275 \\

Sandy Lyle &        231 &      4.693 &      3.195 &      0.022 &      7.910 \\

Scott Simpson &        231 &      5.173 &      3.195 &      0.009 &      8.377 \\

Steve Jurgensen &        231 &      0.077 &      3.195 &     -0.001 &      3.270 \\

Fulton Allem &        232 &      9.626 &     -1.336 &      0.015 &      8.305 \\

Larry Rinker &        232 &      5.058 &      3.195 &      0.016 &      8.269 \\

Brian Henninger &        233 &      8.170 &      3.195 &      0.020 &     11.385 \\

Jim Carter &        233 &      9.331 &      3.195 &      0.023 &     12.549 \\

Joe Durant &        233 &     12.828 &     -2.062 &      0.021 &     10.787 \\

 Rich Beem &        233 &      5.681 &      3.195 &      0.007 &      8.882 \\

Tommy Armour III &        233 &     13.932 &     -1.336 &      0.046 &     12.642 \\

Trevor Dodds &        233 &      6.757 &      3.195 &      0.003 &      9.955 \\

J.P. Hayes &        234 &     14.870 &     -1.859 &      0.030 &     13.041 \\

Cameron Beckman &        237 &     12.456 &     -1.336 &      0.037 &     11.157 \\

Mark Wiebe &        242 &     17.460 &      3.195 &      0.026 &     20.681 \\

\end{tabular}

\end{center}
\end{table}

\clearpage

\begin{table}[loc=h]
\caption{Effects of Playing with Tiger Woods}
\label{id}
\begin{center}

\begin{tabular}{clrr}
\hline
     Test & Dummy variable &      Coef. &    p-value \\
\hline
         1 & Woods in the field &      0.051 &     0.0211 \\

           &            &            &            \\

          2 & Woods in the field &            &            \\

        & \;\;  Playing with Woods &      0.478 &     0.0004 \\

           & \;\;  Not playing with Woods &      0.043 &     0.0505 \\

           &            &            &            \\

        3 & Playing with Woods &      0.462 &     0.0005 \\

           &            &            &            \\

         4 & Playing with Woods &            &            \\

           &  \;\;  Round 1 &      0.611 &     0.0141 \\

           &  \;\;  Round 2 &      0.317 &     0.2006 \\

           &   \;\; Round 3 &      0.063 &     0.8268 \\

           &  \;\;  Round 4 &      0.857 &      0.0030 \\

           &            &            &            \\

         5 & Playing in last scheduled round of tournament &            &            \\

              &  \;\;  Playing with Woods &      0.858 &     0.0030 \\

              &  \;\;  Not playing with Woods &      0.008 &     0.7790 \\

        &            &            &            \\

         6 & Playing with Woods &            &            \\

           &  \;\;  Round 1 &      0.611 &     0.0141 \\

           &  \;\;  Round 2 &      0.317 &     0.2006 \\

           &  \;\;  Round 3 &      0.063 &     0.8268 \\

           &    \;\; Round 4 playing with Woods in final group  &     0.389 &     0.4184 \\

           &   \;\; Round 4 playing with Woods but not in final group &      1.122 &     0.0019 \\

                                    &            &            &            \\

           &    \;\; Round 4 playing with Woods, Tiger within 10 strokes of the lead &     0.796 &     0.0087 \\
            &   \;\; Round 4 playing with Woods, Tiger not within 10 strokes of the lead &      1.460 &     0.1232 \\

                               &            &            &            \\

           &    \;\; Round 4 playing with Woods, Tiger within 8 strokes of the lead &     0.801 &     0.0088 \\
            &   \;\; Round 4 playing with Woods, Tiger not within 8 strokes of the lead &      1.349 &     0.1308 \\

                     &            &            &            \\

           &    \;\; Round 4 playing with Woods, Tiger within 6 strokes of the lead &     0.812 &     0.0118 \\
            &   \;\; Round 4 playing with Woods, Tiger not within 6 strokes of the lead &      1.044 &     0.1083 \\

                                &            &            &            \\

           &    \;\; Round 4 playing with Woods, Tiger within 4 strokes of the lead &     0.878 &     0.0297 \\
           &   \;\; Round 4 playing with Woods, Tiger not within 4 strokes of the lead &      0.837 &     0.0430 \\

                               &            &            &            \\

           &    \;\; Round 4 playing with Woods, Tiger within 2 strokes of the lead &     0.690 &     0.1125 \\
           &   \;\; Round 4 playing with Woods, Tiger not within 2 strokes of the lead &      0.991 &     0.0104 \\

                             &            &            &            \\
          &    \;\; Round 4 playing with Woods, Tiger tied for lead or better &     0.329 &     0.5649 \\

          &   \;\; Round 4 playing with Woods, Tiger not tied for lead or better &      1.040 &     0.0019 \\

\hline
\end{tabular}
\end{center}
\end{table}

\clearpage

\begin{table}[loc=h]
\caption{Effects of Being in Contention in Final Scheduled Round}

\begin{center}

\begin{tabular}{clcc}
\hline

      Test & Dummy variable &      Coef. &    p-value \\

\hline

             1 & Player within 10 strokes of lead &      0.042 &     0.2380 \\

           & Player not within 10 strokes of lead &     -0.036 &     0.3699 \\

           & Playing with Woods &      0.822 &     0.0046 \\

           &            &            &            \\

         2 & Player within 8 strokes of lead &      0.104 &     0.0160 \\

           & Player not within8 strokes of lead &     -0.049 &     0.1514 \\

           & Playing with Woods &      0.775 &     0.0078 \\

           &            &            &            \\

         3 & Player within 6 strokes of lead &      0.159 &     0.0043 \\

           & Player not within 6 strokes of lead &     -0.034 &     0.2775 \\

           & Playing with Woods &      0.743 &     0.0109 \\

           &            &            &            \\

         4 & Player within 4 strokes of lead &      0.171 &     0.0297 \\

           & Player not within 4 strokes of lead &     -0.011 &     0.6997 \\

           & Playing with Woods &      0.780 &     0.0075 \\

           &            &            &            \\

         5 & Player within 2 strokes of lead &      0.185 &     0.1124 \\

           & Player not within 2 strokes of lead &     -0.001 &     0.9788 \\

           & Playing with Woods &      0.794 &     0.0066 \\

           &            &            &            \\

         6 & Player in the lead (or tied) &      0.193 &     0.3115 \\

           & Player not in the lead &      0.005 &     0.8675 \\

           & Playing with Woods &      0.822 &     0.0048 \\

\hline
\end{tabular}

\end{center}
\end{table}

\clearpage

\begin{sidewaystable}[loc=h]
\caption{Phil Mickelson's Performance in Majors, 1998-2001}
\label{id}
\begin{center}

\begin{tabular}{lrrrrrrlrrl}
\hline
           &            &            &     Mickelson's &     Mickelson's &            &            &            &            &            &            \\

           &    Winning &     Mickelson's &   expected &   residual &            & \multicolumn{ 2}{c}{If played like} &            & \multicolumn{ 2}{c}{If played like} \\

           &      score &      score &      score &      score &            & \multicolumn{ 2}{c}{1999 U.S. Open} &            & \multicolumn{ 2}{c}{2001 PGA} \\

           &        (1) &        (2) &        (3) &        (4) &            & \multicolumn{ 2}{c}{(5)} &            & \multicolumn{ 2}{c}{(6)} \\
 \cline{1-5} \cline{7-8} \cline{10-11}

98 Masters &        279 &        286 &     290.66 &      -4.66 &            &     278.35 &        Win &            &     281.17 &       Lose \\

98 U.S. Open &        280 &        288 &     291.59 &      -3.59 &            &     279.28 &        Win &            &     282.10 &       Lose \\

98 British Open &        280 &        308 &     293.38 &      14.62 &            &     281.07 &       Lose &            &     283.89 &       Lose \\

98 PGA &        271 &        285 &     285.04 &      -0.04 &            &     272.73 &       Lose &            &     275.55 &       Lose \\

           &            &            &            &            &            &            &            &            &            &            \\

99 Masters &        280 &        285 &     290.67 &      -5.67 &            &     278.35 &        Win &            &     281.17 &       Lose \\

99 U.S. Open &        279 &        280 &     292.31 &     -12.31 &            &     280.00 &       Lose &            &     282.82 &       Lose \\

99 British Open &                                             \multicolumn{ 6}{c}{Missed Cut} &       Lose &            &            &       Lose \\

99 PGA &        277 &        295 &     288.48 &       6.52 &            &     276.17 &        Win &            &     278.99 &       Lose \\

           &            &            &            &            &            &            &            &            &            &            \\

00 Masters &        278 &        286 &     288.67 &      -2.67 &            &     276.36 &        Win &            &     279.18 &       Lose \\

00 U.S. Open &        272 &        293 &     293.42 &      -0.42 &            &     281.11 &       Lose &            &     283.93 &       Lose \\

00 British Open &        269 &        281 &     280.96 &       0.04 &            &     268.65 &        Win &            &     271.47 &       Lose \\

00 PGA &        270 &        279 &     281.33 &      -2.33 &            &     269.02 &        Win &            &     271.84 &       Lose \\

           &            &            &            &            &            &            &            &            &            &            \\

01 Masters &        272 &        275 &     280.91 &      -5.91 &            &     268.59 &        Win &            &     271.41 &        Win \\

01 U.S. Open &        276 &        282 &     283.48 &      -1.48 &            &     271.17 &        Win &            &     273.99 &        Win \\

01 British Open &        274 &        285 &     280.66 &       4.34 &            &     268.35 &        Win &            &     271.17 &        Win \\

01 PGA &        265 &        266 &     275.49 &      -9.49 &            &     263.18 &        Win &            &     266.00 &     Lose \\
\hline
\end{tabular}

\end{center}
\end{sidewaystable}

\clearpage

\begin{table}[loc=h]
\caption{Mean $\theta$ Residual by Round in Majors, 1998-2001}

\label{id}
\begin{center}

\begin{tabular}{lrrrr}
\hline
           &            &      Woods &            &       Mickelson \\

  \cline{3-3} \cline{5-5}
   Round 1 &            &     -0.632 &            &     -0.903 \\

   Round 2 &            &     -0.610 &            &     -1.421 \\

   Round 3 &            &     -0.042 &            &      0.651 \\

   Round 4 &            &     -0.244 &            &      0.519 \\
  \cline{3-3} \cline{5-5}
           &            &            &            &            \\

   Overall &            &     -0.382 &            &     -0.317 \\

           &            &            &            &            \\

 Std error &            &      0.297 &            &      0.352 \\
\hline
\end{tabular}
\end{center}
\end{table}

\end{document}